\documentclass[%
 reprint,
 amsmath,amssymb,
 aps,
 prl,
]{revtex4-1}

\usepackage{graphicx}
\usepackage{dcolumn}
\usepackage{bm}
\usepackage{color}

\begin{document}

\preprint{APS/123-QED}
\title{Asymmetric acoustic propagation of wave packets via the self-demodulation effect }

\author{Thibaut Devaux}
  \email{To whom correspondence should be addressed.\\thibaut.devaux@univ-lemans.fr}
\author{Vincent Tournat}%


\author{Olivier Richoux}
\author{Vincent Pagneux}
\affiliation{%
 LUNAM Universit\'{e}, Universit\'{e} du Maine, CNRS, LAUM UMR 6613, avenue O. Messiaen, 72085 Le Mans, France.
}%


\date{\today}

\begin{abstract}
This article presents the experimental characterization of nonreciprocal elastic wave transmission in a single-mode elastic
waveguide. 
This asymmetric system is obtained by coupling a selection layer with a conversion layer: 
the selection component is provided by a phononic crystal, while the conversion is achieved by a nonlinear self-demodulation effect
in a  3D unconsolidated granular medium. 
A quantitative experimental study of this acoustic rectifier indicates a high rectifying ratio, up to $10^6$, with 
wide band ($10$ kHz) and an audible effect. 
Moreover, this system allows for wave-packet  rectification and extends the future applications of asymmetric systems.

\end{abstract}

\maketitle


Over the past decade, the development of asymmetric systems operating on acoustic waves has proven to be a real challenge given the numerous applications both in optics and for radio-waves \cite{what_is_opt}. Research on unidirectional transmission devices for acoustic and elastic waves, i.e. permitting the wave energy to pass through in one direction but not the other, has led to applications, such as energy control, energy harvesting or accumulation, the transistor effect, logic gates and the memory effect for thermal devices \cite{diode_th_lattices,transistor_th2,logic_gate_th,memory_th}, as well as to operations on signals and data, e.g. the optical device proposed in \cite{diode_opt_heterojunctions,Wang_PRL_2013,diode_opt_periodic}.
These advances have been able to dramatically improve: comfort in noisy environments, the stealth of noisy or sound-reflecting objects, and the quality of non-destructive testing or medical imaging with ultrasound \cite{nature_revolution,li2010now,diode_waveguide}. Moreover, they have made positive contributions to features like shock protection and interface identification \cite{diode_granulaire}. 

It has been clearly recalled in recent articles that to obtain an asymmetric transmission of waves capable of replicating the effect of an isolator, the reciprocity of a wave needs to be broken \cite{diode_what_is,what_is_opt}. To achieve this reciprocity break, most widely known solutions consist of using nonlinearity, breaking the time invariance of the system by modulating some of its properties over time, or else biasing the system with a vectorial field \cite{diode_circulator,PREMerkel} whose effect differs between forward and backward propagation (e.g. the magnetic field for the Faraday isolator \cite{what_is_opt}). In acoustics, asymmetric systems relying on nonlinearity have been based, for example, on the second harmonic generation \cite{diode_bulle,diode_multi} or  the bifurcation process \cite{diode_bille}. In both cases, the system is composed of a nonlinear medium (bubbly water) \cite{diode_bulle} or localized nonlinear element (a nonlinear defect in  the granular chain) \cite{diode_bille}, thus converting the acoustic energy into different frequencies than the emitted one. Futhermore, a selection layer opaque to the initially emitted frequencies is laid out on one side of the nonlinear frequency converter; this might consist of a phononic crystal with a forbidden band gap. In one  propagation direction, the emitted wave impinges the selection layer and is nearly totally reflected. In the other direction, the emitted frequencies are altered in the nonlinear medium and, provided an appropriate design,  transmitted through the selection layer. 

An important parameter in quantifying the transmission asymmetry of this system is the asymmetry ratio $\sigma$ (also denoted as the rectification ratio in \cite{diode_bulle}), which is defined as the ratio of the energy transmission coefficients through the system in both propagation directions. In \cite{diode_bulle}, an asymmetry ratio of $\sigma \approx 10^4$ was experimentally obtained with an overall weak energy transmission coefficient of $\sim 10^{-3}$. The device in \cite{diode_bille} combined a $\sigma \approx 10^4$ asymmetry ratio with a $\sim 35$\% energy transmission value. An air flow-induced bias has been implemented in \cite{Estep_2014,Sounas2013} in order to break the wave transmission reciprocity among the three acoustic inputs and outputs of a circulator. This design, based on three single-mode waveguides coupled to a cavity where a circulating air flow is present (at the cost of external energy), exhibits an isolation of $40$ dB between two ports, while the transmission value lies close to 1 in the other propagation direction. Moreover, unlike devices that involve nonlinearity, the frequency content between input and output waves in this circulator remains the same. Unfortunately, in all of these nonlinear or biased devices, the nonreciprocal characteristic is present within a narrow frequency range, basically only allowing operations with continuous waves. One of the limitations inherent in the previously proposed acoustic asymmetric systems is an inability to combine a high level of asymmetry with a wide operating band. An active electro-acoustic membrane coupled with Helmholtz resonators and connected to an appropriately designed nonlinear electrical circuit, constituted a nonreciprocal element implemented in \cite{diode_piezo} for building an asymmetric system operating with acoustic wave packets in air. This time domain operation with wave packets has provided an opportunity to introduce information coding capabilities, which may potentially  be very attractive in many applications.

In this letter, we are proposing the design and operations of an elastic wave rectifier   based on the nonlinear self-demodulation effect. Such a device offers a high rectification ratio (up to $10^6$) and a wide operating band  ($> 80 \%$, for a ratio above $10^5$). It should be pointed out that the proposed device architecture does not require a precise design of the acoustic filter due to the frequency-down conversion of the self-demodulation effect, requiring transparency over a low-frequency range and opaqueness for a higher-frequency band, which in practice is easy to achieve. In addition, the rectifier is to operate with elastic waves in solids without requiring any external energy.

\begin{figure}[htbp]
\centering
\includegraphics[width= 0.48\textwidth]{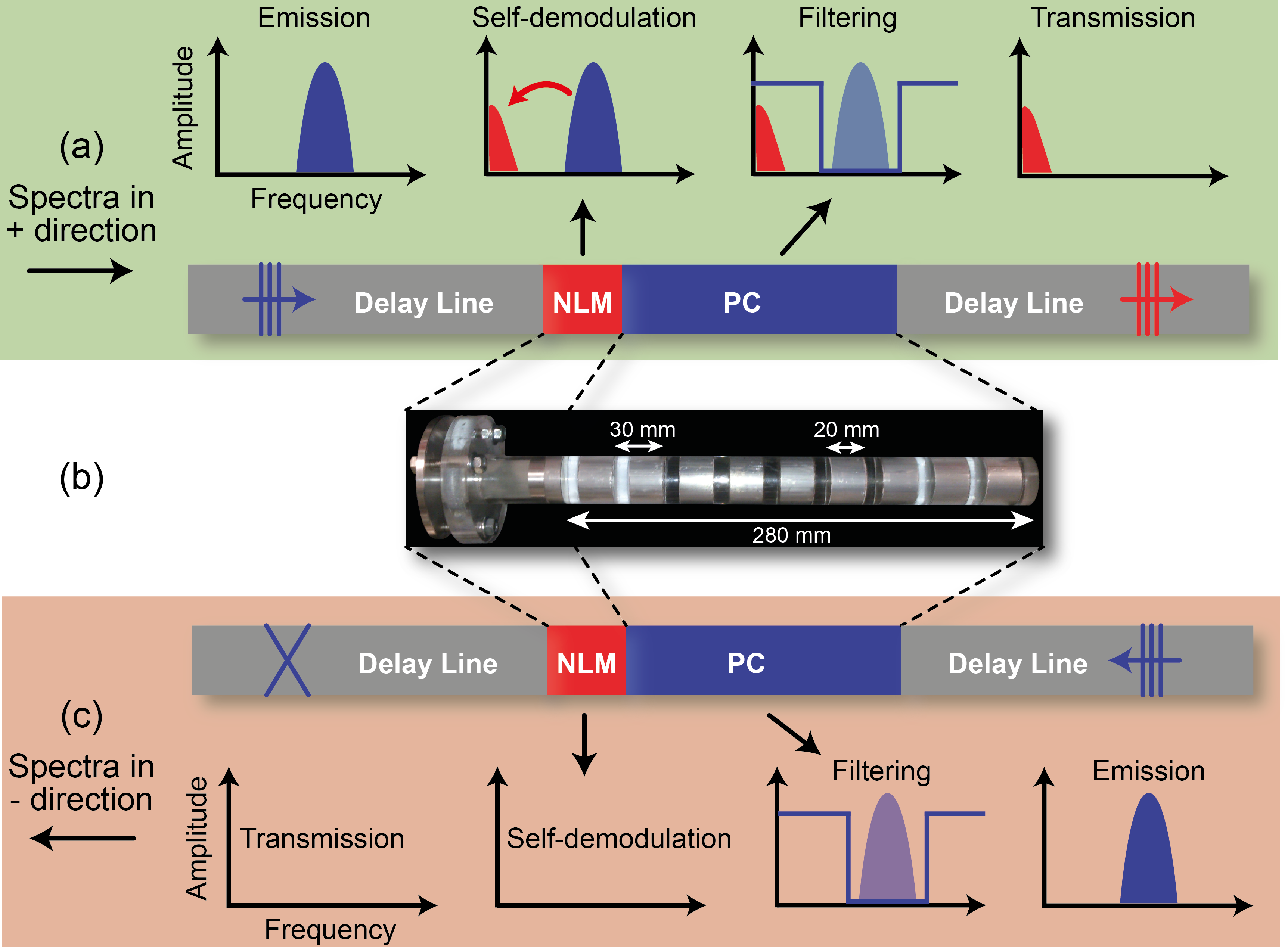}
\caption{\label{fig_principe} (a) Schematics of the rectifier for the "$+$" direction;  (b) Picture of the rectifier core, composed of a multilayer PC connected to a granular layer; and  (c) Schematics for the "$-$" direction.
}
\normalsize
\end{figure}

The schematic principle and a picture of the rectifier are shown in Fig.~\ref{fig_principe}. The selection layer is a periodic medium, i.e.  a phononic crystal (PC), composed of 9 alternating layers of aluminum and lucite. The conversion is performed by a strongly nonlinear granular medium  (NLM) through a nonlinear self-demodulation effect.

Let us specify the "$-$" direction for the case where the emitted wave, launched from the right side, first  meets the PC, as shown in Fig.~\ref{fig_principe}\textbf{(c)}.
In this direction, an amplitude-modulated wave first excites the PC, which has been designed to exhibit a passband below a cutoff frequency of $f_c \simeq 35 kHz$ and a band gap above. If the excitation signal frequencies exceed $f_c$, then: the PC does not transmit the signal, no acoustic energy is conveyed to the granular medium, and no signal passes through the system.

In contrast,  in the "$+$" direction, see Fig.~\ref{fig_principe}(a), the wave, launched from the left side,  first excites the nonlinear granular medium, which converts part of the initial acoustic energy to lower frequencies through the self-demodulation effect (a frequency-down conversion by means of difference-frequency generation) \cite{granular_demodPRL}. At the interface between the granular medium and the multilayer PC, the initial signal and the nonlinearly self-demodulated signal are both present. The PC then filters out frequencies above $f_c$ while still transmitting the frequency components below $f_c$: the signal initially emitted is reflected, but the self-demodulated signal located in a passband is transmitted. 

In comparing the design of this architecture to the rectifier proposed by B. Liang \textit{et al.} \cite{diode_bulle}, the frequency-down conversion with self-demodulation requires a simpler design than the frequency-up conversion with the second harmonic generation, because, in general a PC is  naturally transparent at sufficiently low frequencies.

\begin{figure}[h!]
	\centering
\includegraphics[width= 0.4\textwidth]{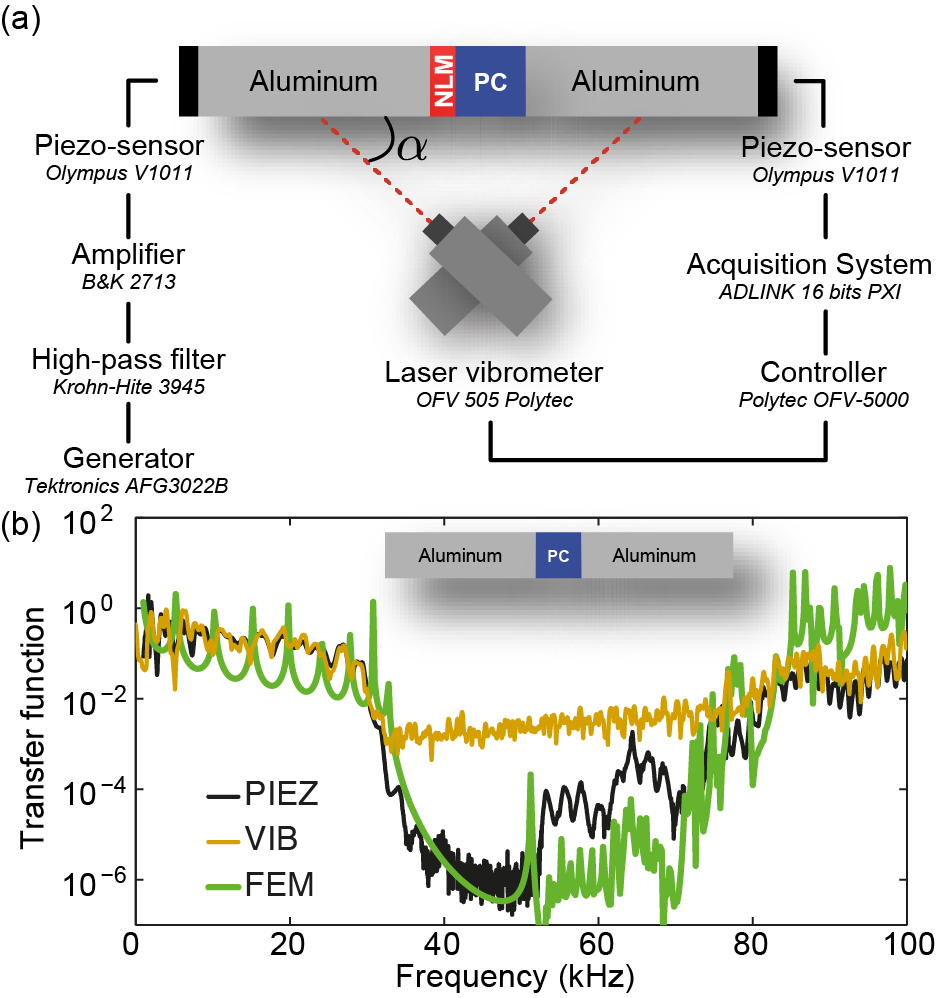}
\caption{\label{fig_setup} (a) Schematic diagram of the experimental set-up in the "+" direction. Electrical connections to the piezo-sensors have been inverted for the "-" direction. (b) Comparison of transfer functions between the simulated type (FEM) and the experimental type (VIB, measured according to the laser vibrometer method and PIEZ, measured with the piezoelectric sensor).
}
\normalsize
\end{figure}

The experimental set-up is presented in Fig. \ref{fig_setup}(a). The acoustic rectifier has been placed between two delay lines, in accordance with the notion inherent in the Hopkinson bar experiment \cite{hopkinson}.
This particular set-up provides a clear determination of the incident, reflected and transmitted pulsed waves, which constitute an important aspect to conducting an unambiguously study of asymmetric transmission.
Delay lines consist of aluminum rods with a circular cross-section, a diameter $d_b=30$ mm and length $L=1$ m, thus ensuring that only a single axisymmetric longitudinal mode is able to propagate at up to ~90 kHz. This single-mode signal is measured in the middle of delay lines with a laser vibrometer at angles $\alpha=38$° and $-\alpha$ relative to the rod axis, making it possible to capture the wave particle velocity both quantitatively and non-invasively. In the general case, this velocity is the projection along the laser beam of an in-plane and out-of-plane motion \cite{footnote}. In addition, two identical piezo-sensors have been glued at both ends of the delay lines; they play the role of source or receiver depending on the selected direction of operations. To avoid possible nonlinear effects from the generator, the input signal is high-pass filtered and  then amplified.

The phononic crystal (PC selection layer)  is composed of $10$ layers of aluminum, alternating with $9$ layers of lucite whose thicknesses are $d_a = 20$ mm and $d_p= 10$ mm, respectively, along with diameters of $30$ mm. This crystal has been assembled along with an epoxy glue. Figure \ref{fig_setup}(b) shows the transfer functions of the PC measured by two experimental methods and computed by Finite Element Method (FEM), without losses. As regards the laser vibrometer method, a wide-band pulse is emitted with the piezo-transducer, in order to identify, within the time domain, the incident and transmitted waves in the delay lines on both sides of the PC by itself. The transfer function is the ratio of the Fourier transform magnitudes of the incoming and outgoing particle velocity pulses. With a second piezo-transducer serving as a receiver, the input signal is a frequency-swept sine, which in conjunction with use of a spectrum analyzer yields a transfer function with a higher Signal-to-Noise Ratio (SNR) than the laser vibrometer method. The FEM transfer function is computed for the PC alone as the ratio between the output and input axial displacements, respectively detected and imposed at both ends, on the PC axis. The experimental methods and numerical simulation results are in good agreement with one another, yet a significant difference in the experimental amplitudes appears  inside the band gap ($[30-80]$ kHz) of this PC, which is due to the relatively lower SNR associated with laser vibrometer method \cite{footnote0,achenbach}.

The nonlinear medium  (NLM conversion layer)  is of the unconsolidated granular type and has been made of steel powder (\textit{Fe3Cr1Mo0.5Mn0.2V}) whereby the various grain sizes average  $d=300$ $\mu m$ in diameter. The powder is confined with a static pressure of $\sim10$ kPa in a lucite tube $h=30$ mm long, with an inner diameter $d_g=30$ mm closed by aluminum discs and coupled with the delay line and the  glued PC. The preferred characteristics of this  granular medium stem from a compromise between low losses and high nonlinearity, in order to maximize the self-demodulation effect for a pulsed signal centered at $45$ kHz with a $10$ kHz bandwidth.

To operate this acoustic rectifier, a wave packet with a carrier frequency $f_0=45$ kHz and characteristic Gaussian envelope duration $t_g=1/f_g=0.1$ ms is excited in the emission delay line. With these parameters, the input signal spectrum  (displayed in Fig.~\ref{fig_freq}(a)) corresponds to the smallest transmission through the PC, Fig.~2(b). In the "+" direction, i.e. Fig. 3(a,b), the excited wave packet first meets  the nonlinear medium and is partially demodulated; its initial spectrum components are then reflected by the PC, Fig.~3(a). The self-demodulated part is transmitted through the PC and reaches the transmission delay line, as observed in Fig.~3(b). Until the time indicated by the vertical dashed line, the signal corresponds to the outgoing wave before reflection by the delay line boundary \cite{footnote2,gran_Norris_1997,gran_johnson1999nonlinear,gran_VT_antenna,granular_self_demodulation_1D}. In the "-" direction, Fig. 3(c,d), the excited wave packet  first meets the PC and is totally reflected (see the incident and reflected signals in Fig.~3(c)). In the transmission delay line, Fig.~3(d), no detectable signal is visible. The asymmetric nature of the rectifier has thus been demonstrated here in the time domain for pulsed signals.

\begin{figure}[t]
	\centering
\includegraphics[width= 0.5\textwidth]{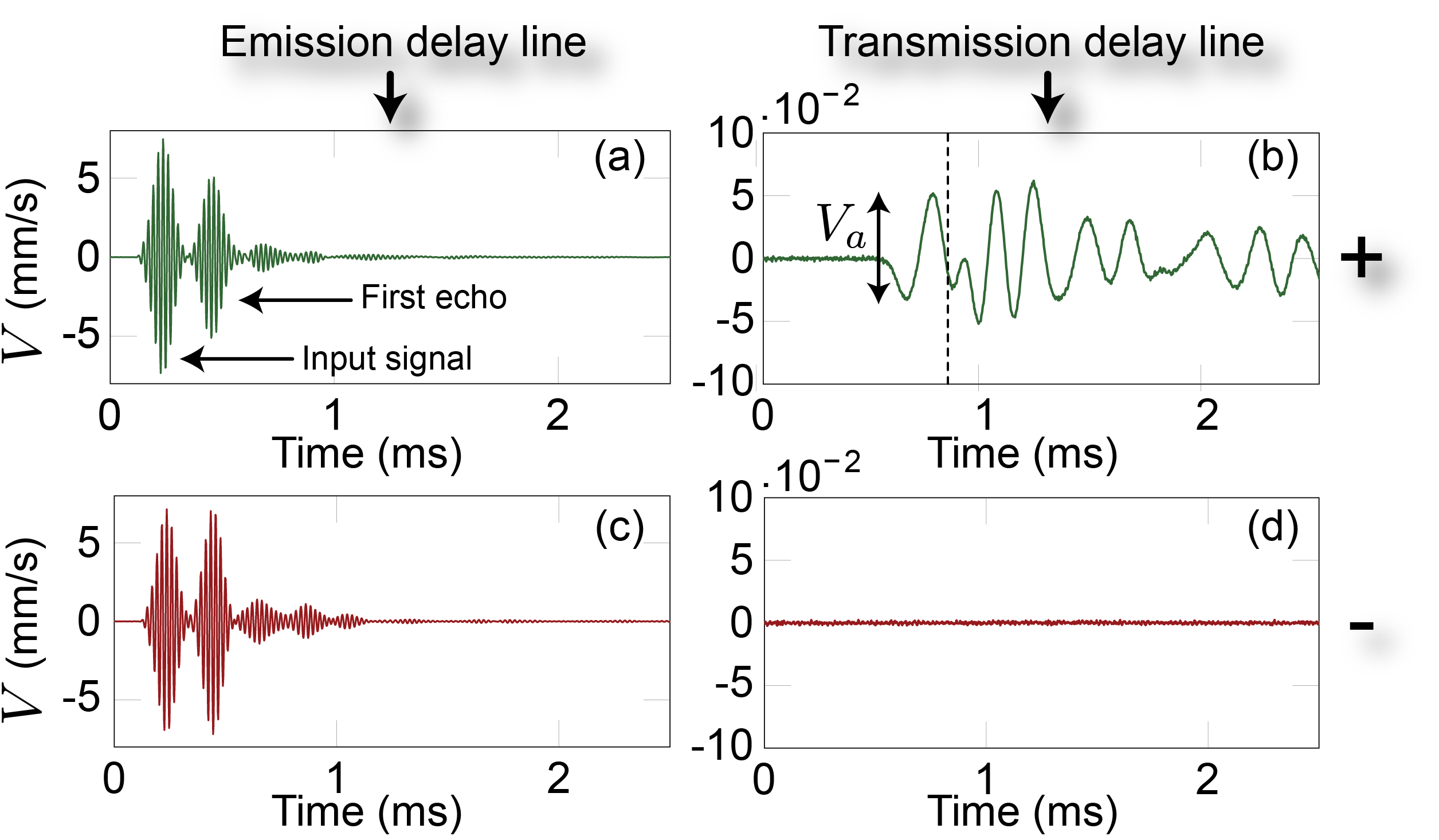}
\caption{\label{fig_temp}Particle velocity $V$ measured by the laser vibrometer in the middle of the delay lines for the largest excitation amplitude. (a) and (b) "+" direction configuration in the emission and transmission delay lines, respectively. (c) and (d) "-" direction configuration in the emission and transmission delay lines, respectively.}
\normalsize
\end{figure}

To fully characterize the rectifier, we must now make use of  information from the piezo-sensors, in generating less background noise than the vibrometer, which becomes a crucial limiting factor notably for determining the transmitted signal amplitudes in the "-" direction. In Fig.~4(a), the squared spectral magnitudes $A^2_s$ of the signals received by the piezo-sensors are plotted for both directions "+" ($A^2_s +$) and "-" ($A^2_s -$). Before calculating the Fourier transform, the piezo-sensor temporal signal $S$ is windowed in order to select the first arrivals and avoid contributions from the multiple reflections in the delay line, as depicted in Fig.~4(b).

\begin{figure}[h]
	\centering
	\includegraphics[width= 0.4\textwidth]{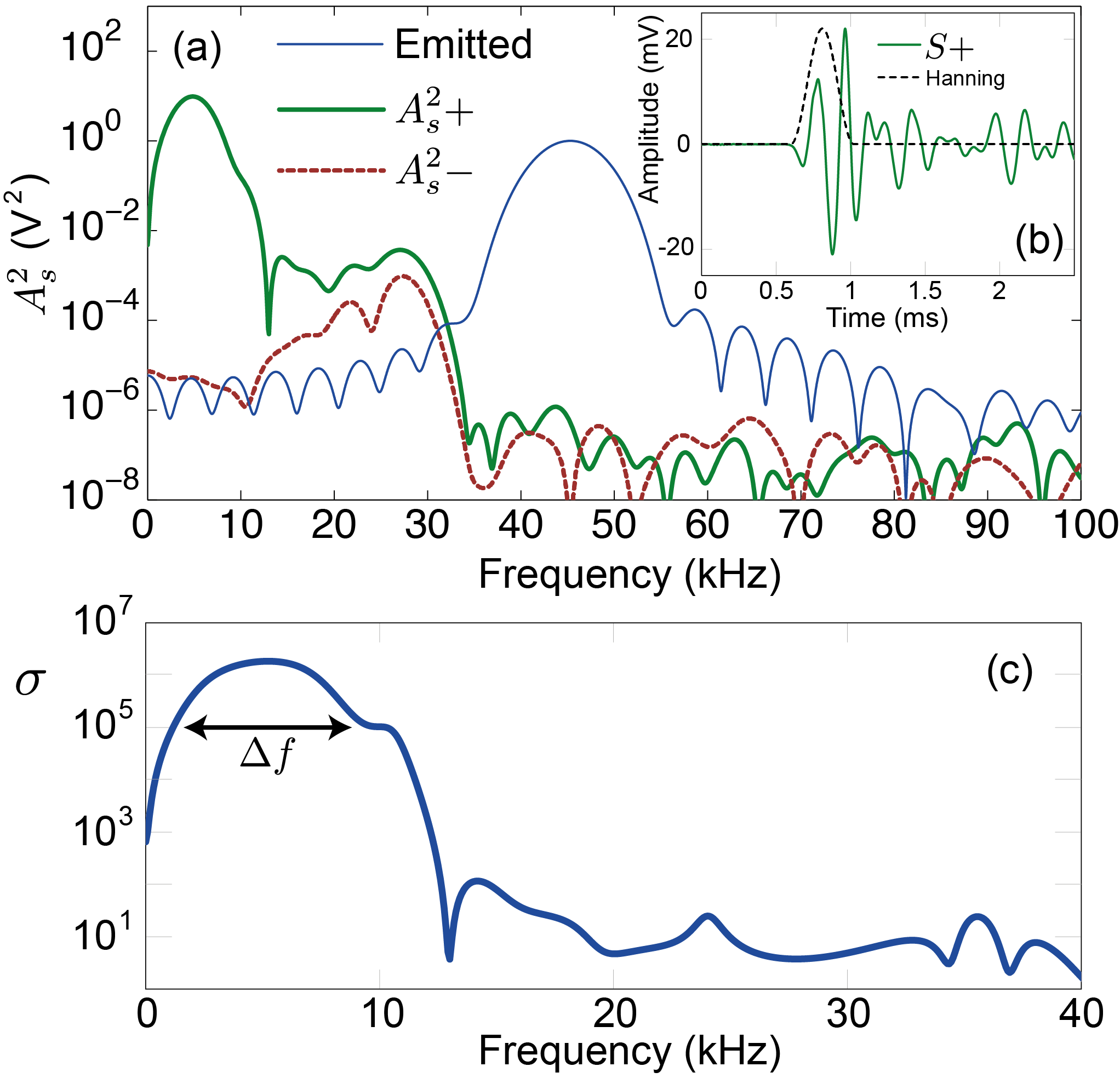}
      \caption{\label{fig_freq} (a) Squared spectral magnitude, transmitted through the rectifier in both the "+" ($A^2_s +$) and "-" ($A^2_s -$) directions, and received by the piezo-sensors.  The normalized squared spectral magnitude of the incident particle velocity wave packet is also displayed. (b) A temporal signal from the piezo-sensor and the Hanning window used for the Fourier transform. (c) Energy rectification ratio of the device, defined as $\sigma=(A^2_s +)/(A^2_s -)$.}
\end{figure}

The rectifier  asymmetry is visible in the frequency domain through the difference between transmitted wave energies in the "+" and "-" directions, see Fig.~4(a) for frequencies less than 10 kHz. To quantify this asymmetry, a frequency-dependent rectification ratio is introduced \cite{diode_bulle}, $\sigma=(A^2_s +)/(A^2_s -)$, and plotted in Fig.~4(c). Let us note that the quantity $A^2_s$ is proportional to the transmitted  elastic wave energy; consequently, $\sigma$ is an energy rectification ratio. In Fig.~4(c), $\sigma$ reaches a maximum above $10^6$, which constitutes a stronger asymmetry than previously reported works \cite{diode_bulle,diode_bille}. Moreover, this value in our experiment is limited by the noise level of the "-" direction signal and could probably be further increased. Over a frequency band $\Delta f \simeq 10$ kHz centered on $\sim 6$ kHz (i.e. a $>80$\% bandwidth), a rectification ratio $\sigma \geq 10^5$ is observed, thus demonstrating the broadband operations of the rectifier. This broadband characteristic may be very important in practice for both information transmission and operational flexibility, in contrast with \cite{diode_bulle}.

\begin{figure}[h]
	\centering
	\includegraphics[width= 0.4\textwidth]{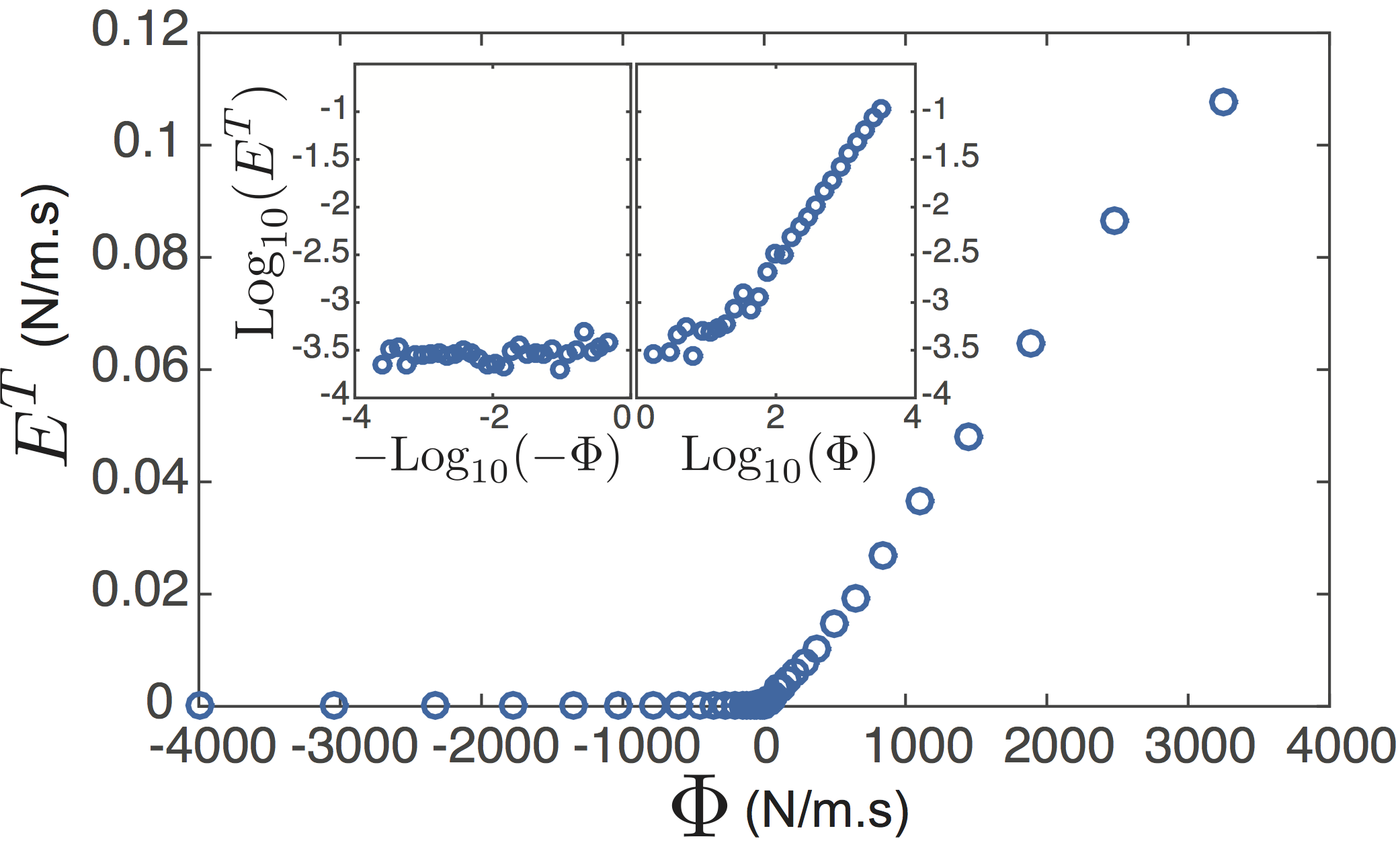}
      \caption{\label{fig_amp}
      Influence of excitation amplitude (vibrometer) on the maximum reception amplitude (vibrometer) for both the $"-"$ and $"+"$ directions,  with an emitted signal composed of a sinus of frequency $f_0=45.0$ kHz modulated in amplitude by a Gaussian envelope of frequency $f_g=10.0$ kHz.
}
\end{figure}

Let us  now examine the amplitude dependence of the rectifier. To perform this step quantitatively, the peak-to-peak amplitude of wave particle velocity $V_a$ is measured using the laser vibrometer in the middle of the emission and transmission delay lines and then time-windowed in order to extract the incident and transmitted wave packets. The amplitude-dependent behavior of the rectifier is presented in Fig.~5 in the form of the transmitted elastic energy $E^T=  E (V_a^{t})^2/c_0$ vs. the differential incident elastic energy $\Phi$, defined as:
\begin{equation}\label{refeqddp}
\Phi = E \dfrac{(V_a^{i+})^2}{c_0}-E \dfrac{(V_a^{i-})^2}{c_0}
\end{equation}
where $E=69$ GPa is the Young's modulus of aluminum, $V_a^{i\pm}$  the peak-to-peak amplitude of the incident wave particle velocity measured in the "+" or "-" direction (first wave packet of Fig.~3(a,c)), $V_a^{t}$  the amplitude of the particle velocity time-windowed to extract solely the outgoing contribution (as in Fig.~3b before the vertical dashed line) and $c_0=4715$ m/s  the wave velocity in the aluminum rod. According to our experimental protocol, we have excited the wave from either one side or the other so that, in each case, one of the terms from Eq.~(\ref{refeqddp}) equals zero. Keep in mind that the quantity $\Phi$ is only roughly estimated here, because of the fact that the elastic energy expression $E (V_a^{i\pm})^2/c_0$ is valid for a one-dimensional motion, even though longitudinal and  radial motions occur in the single axisymmetric propagation mode at 45 kHz. Also, the particle velocity $V_a^{i\pm}$ has been detected  as a combination of these two contributions  \cite{footnote}. 
On the other hand, the transmitted elastic energy $E^{T}$ is better defined and more accurately measured since for the lower associated frequencies $\sim 5$ kHz, the motion is mainly one-dimensional and dominated by the longitudinal displacement \cite{achenbach}. Therefore, the particle velocity amplitude $V_a^{t}$, corresponds to the longitudinal motion. The device characteristics in Fig.~5 depict its asymmetry  and  amplitude dependence. For negative values of $\Phi$, the transmitted elastic energy $E^T$ is close to zero and limited by the experimental system noise (see the Log-Log plot in the inset). For positive values of $\Phi$, the amplitude dynamics seem to be essentially linear $E^T \propto \Phi$, in the main panel of Fig.~5. For small positive values of $\Phi$ (typically below 200 N/m.s) however, the amplitude dynamics differs and could be approximated locally by a power law. This behavior originates from the amplitude dynamics of the self-demodulation effect in the nonlinear granular powder layer placed in the device. At vanishing excitation amplitudes, the self-demodulated energy is shown to be quadratic in excitation energy \cite{granular_demodPRL}. For higher excitation amplitudes, a 3/2 power law, due to the appearance of the clapping phenomenon between contacts, has been reported, in saturating towards a near linear trend line (power 1) for higher amplitudes where nonlinear dissipation is expected to exert influence \cite{granular_demodPRL}. 

In conclusion, a new asymmetric wave  transmitter architecture has been proposed by utilizing physical conversion and selection mechanisms. The nonlinear self-demodulation effect induced by unconsolidated granular media has been introduced for the conversion process, while the selection procedure has been completed with a phononic crystal. Experimental results indicate a high contrast ratio $\sigma \approx 10^6$, associated with a low-frequency broad range $\Delta f \approx 10$ kHz. The amplitude-dependent behavior of the proposed device exhibits similarities with electrical diodes and leads to new applications for wave control through asymmetric systems.

\begin{acknowledgments}
This work has been financed by "DGA" and  the "PROPASYM" project funded by the R\'{e}gion Pays-de-la-Loire.
\end{acknowledgments}


\begin{thebibliography}{10}


\bibitem{what_is_opt}
D. Jalas, A. Petrov, M. Eich, W. Freude, S. Fan,
  Z. Yu, R. Baets, M. Popovic, A. Melloni, J.~D. Joannopoulos,
  et~al,
\newblock {Nat. Photonics} \textbf{7}, 579 (2013).


\bibitem{diode_th_lattices}
B. Li, J.H. Lan, and L. Wang,
\newblock {Phys. Rev. Lett.} \textbf{95}, 104302 (2005).

\bibitem{transistor_th2}
B. Li, L. Wang, and G. Casati,
\newblock {Appl. Phys. Lett.} \textbf{88}, 143501 (2006).

\bibitem{logic_gate_th}
L. Wang and B. Li,
\newblock {Phys. Rev. Lett.}, \textbf{99}, 177208 (2007).

\bibitem{memory_th}
L. Wang and B. Li,
\newblock { Phys. Rev. Lett.} \textbf{101}, 267203 (2008).

\bibitem{diode_opt_heterojunctions}
J.~{Hwang}, M.~H. {Song}, B.~{Park}, S.~{Nishimura}, T.~{Toyooka}, J.~W. {Wu},
  Y.~{Takanishi}, K.~{Ishikawa}, and H.~{Takezoe}.
\newblock {Nat. Mater.} \textbf{4}, 383 (2005).

\bibitem{Wang_PRL_2013}
D.-W. Wang, H.-T. Zhou, M.-J. Guo, J.-X. Zhang, J. Evers, and
  S.-Y. Zhu,
\newblock {Phys. Rev. Lett.} \textbf{110}, 093901 (2013).

\bibitem{diode_opt_periodic}
K. Gallo, G. Assanto, K.R. Parameswaran, and M.M. Fejer,
\newblock { Appl. Phys. Lett.} \textbf{79}, 314 (2001).

\bibitem{nature_revolution}
M. Maldovan,
\newblock {Nature} \textbf{503}, 209 (2013).

\bibitem{li2010now}
B. Li.
\newblock {Nat. Mater.} \textbf{9}, 962 (2010).

\bibitem{diode_waveguide}
B. Yuan, B. Liang, J. C. Tao, X.Y. Zou, and J.C. Cheng,
\newblock { Appl. Phys. Lett.} \textbf{101}, 043503 (2012).

\bibitem{diode_granulaire}
V.F.~Nesterenko, C. Daraio, E.B.~Herbold, and S. Jin,
\newblock {Phys. Rev. Lett.} \textbf{95}, 158702 (2005).



\bibitem{diode_what_is}
A.~A. {Maznev}, A.~G. {Every}, and O.~B. {Wright},
\newblock {Wave Motion} \textbf{50}, 776 (2013).


\bibitem{PREMerkel}
A. Merkel, V. Tournat, and V.E. Gusev,
\newblock { Phys. Rev. E} \textbf{90}, 023206  (2014).


\bibitem{diode_circulator}
R. Fleury, D.L. Sounas, C.F. Sieck, M.R. Haberman, and A. Al{\`u},
\newblock {Science} \textbf{343}, 516 (2014).


\bibitem{diode_bulle}
B.~Liang, B.~Guo, J.~Tu, D.~Zhang, and J.~C. Cheng,
\newblock {Nat. Mater.} \textbf{9}, 989 (2010).

\bibitem{diode_multi}
B.~Liang, B.~Yuan, and J.C. Cheng,
\newblock {Phys. Rev. Lett.} \textbf{103}, 104301 (2009).

\bibitem{diode_bille}
N.~Boechler, G.~Theocharis, and C.~Daraio,
\newblock {Nat. Mater.} \textbf{10}, 665 (2011).


\bibitem{Estep_2014}
N.A. Estep, D.L. Sounas, J. Soric, and A. Alu,
\newblock {Nat. Phys.} \textbf{10}, 923 (2014).


\bibitem{Sounas2013}
D.L. Sounas, C. Caloz, and A.~Al{\`u},
\newblock {Nat. Commun.} \textbf{4}, 2407 (2013).

%
%

\bibitem{diode_piezo}
B-I. Popa and S.A. Cummer,
\newblock {Nat. Com.} \textbf{5}, 3398 (2014).



\bibitem{granular_demodPRL}
V.~Tournat, V.~Zaitsev, V.~Gusev, V.~Nazarov, P.~B\'equin, and B.~Castagn\`ede,
\newblock {Phys. Rev. Lett.} \textbf{ 92}, 085502 (2004).


\bibitem{hopkinson}
B.~A. Gama, S.~L. Lopatnikov, and J.~W. Gillespie, Jr,
\newblock {Applied Mechanics Reviews} \textbf{57}, 223  (2004).


\bibitem{footnote}
{The relationship between  longitudinal and radial displacements can be extracted from the Pochhammer-Chree Equation \cite{achenbach} for a compressional wave in rods, i.e.: $\frac{u_r}{u_z}= \frac{p J_1(pa)+ i k C' J_1(qa)}{i k J_0(pa) + q C' J_0(qa)} $ with $C'= \frac{-2 i k p J_1(pa) }{(q^2-k^2) J_1(qa)^2}$, $p= \sqrt{\frac{\omega^2}{c_L^2}-k^2} $, $q= \sqrt{\frac{\omega^2}{c_T^2}-k^2} $, $u_r$: radial displacement, $u_z$: longitudinal displacement, $a$: rod radius. For this set-up with an aluminum rod, a numerical application yields,  at 10 kHz: $\frac{u_r}{u_z} \approx 0.06  $ and at 45 kHz: $\frac{u_r}{u_z} \approx 0.34$.}

%
%
%
%
\bibitem{footnote0}
As estimated from the Pochammer-Chree theory \cite{achenbach}, the second axisymmetric mode in the PC lucite rods propagates above $\sim$50 kHz, which manifests in Fig.~\ref{fig_setup}(b) by the several peaks in the band-gap above this frequency, captured by the FEM analysis.

\bibitem{achenbach}
J. Achenbach,
\newblock {\em Wave propagation in elastic solids}. North-Holland series in applied mathematics and mechanics. North-Holland Pub. Co. (1973).

%


\bibitem{footnote2}

The profile of the transmitted demodulated velocity pulse is close to the first derivative of a Gaussian function (the intensity envelop shape of the primarily emitted wave packet), as expected from the ordinary third order elasticity theory applied to granular media \cite{gran_Norris_1997,gran_johnson1999nonlinear,gran_VT_antenna,granular_self_demodulation_1D}.
 


\bibitem{gran_Norris_1997}
A.~N. Norris and D.~L. Johnson,
\newblock {J.  App. Mech.} \textbf{64}, 39 (1997).

\bibitem{gran_johnson1999nonlinear}
D.~L. Johnson.
\newblock {J. Acoust. Soc. Am.} \textbf{105}, 3087 (1999).

\bibitem{gran_VT_antenna}
V.~Tournat, V.~E. Gusev, and B.~Castagn\`ede,
\newblock {Phys. Rev. E} \textbf{66}, 041303 (2002).



\bibitem{granular_self_demodulation_1D}
V.~Tournat, V.~E. Gusev, and B.~Castagn\`ede,
\newblock {Phys. Rev. E} \textbf{70}, 056603 (2004).



\end{thebibliography}


\end{document}